\documentclass[showpacs,preprintnumbers,amsmath,amssymb]{revtex4}
\usepackage[dvips]{graphicx}
\usepackage{comment}
\begin{document}
\title{Long range absorption in the  scattering of  $^{6}$He 
on $^{208}$Pb and $^{197}$Au at 27 MeV.}

\author{O. R. Kakuee$^{a,c}$,  M. A. G. Alvarez$^{c}$,M. V. Andr\'{e}s$^{c}$,
 S. Cherubini$^{d,f}$, T. Davinson$^{b}$, A. Di Pietro$^b$, W.
Galster$^{d}$, J. G\'{o}mez-Camacho$^{c}$, A. M. Laird$^{b}$, 
M. Lamehi-Rachti$^{a}$, I. Martel$^{e}$, A. M. Moro$^{c}$,
J. Rahighi$^{a,b}$, A. M. S\'{a}nchez-Benitez$^{e}$,
 A. C. Shotter$^{b}$, W. B. Smith$^{b}$, J. Vervier$^{d}$, P. J. Woods$^{b}$
.}

\affiliation{$^{a}$Van De Graaf Laboratory, Nuclear Research Centre, AEOI, P. O. Box 
14155-1339, Tehran, Iran}
\affiliation{$^{b}$Department of Physics and Astronomy, Edinburgh University EH9 3JZ, UK}
\affiliation{$^{c}$Departamento de Física Atómica, Molecular y Nuclear, 
Apartado 1065, Universidad de Sevilla, E-41080 Sevilla, Spain}
\affiliation{$^{d}$Department de Physique Nucleaire, Universite Catholique, Louvain-la-Neuve, Belgium}
\affiliation{$^{e}$Departamento de Física Aplicada, Universidad de Huelva, E-21819, Huelva, Spain}
\affiliation{$^{f}$Institut fur Experimentalphysik III, Ruhr-Universitat Bochun, Germany.} 

\begin{abstract}
\noindent
Quasi-elastic scattering of $^{6}$He at $\it {E}_{lab}$=27 MeV from $^{197}$Au has been
measured in the angular range of $6^\circ-72^\circ$ in the laboratory system
employing LEDA and LAMP detection systems. These data, along previously 
analysed data of 
$^{6}$He + $^{208}$Pb at the same energy, are analysed using Optical Model calculations. 
The role of Coulomb dipole polarizability has been investigated. 
Large imaginary diffuseness parameters are required to fit the data.
This result is an evidence for long range absorption mechanisms in  $^{6}$He
induced  reactions.

\bigskip


\end{abstract}

\date{\today}
\maketitle

\section{Introduction}
 The $^{6}$He nucleus has a weakly bound three-body Borromean n-n-$\alpha$ 
structure 
and is known to have an extended two neutron distribution ~\cite{Tani96,Hansen87}.
Scattering of $^{6}$He is therefore of considerable interest in nuclear physics
since experiments may demonstrate sensitivity to this underlying exotic structure.
 
The interest of measuring elastic scattering of $^6$He arises from the weakly bound nature 
of the projectile, which affects the dynamics of the collision.
$^6$He is bound by less than 1 MeV,  so we expect that the coupling to 
the continuum can significantly modify cross sections in the elastic channel.
Korsheninnikov et al.~\cite{Korsh97}, reported on the measurement of proton
elastic scattering by $^{6}$He, $^{8}$He and $^{11}$Li.

One important aspect that we want to investigate is the role of Coulomb
dipole polarizability \cite{May94,May95}, which consists in the effect 
on the elastic channel of coupling to the break-up channels produced by the
Coulomb dipole force. This effect has been investigated for the
scattering of other weakly bound nuclei on heavy targets, such as d+ $^{208}$Pb \cite{Moro99}, 
$^{7}$Li + $^{208}$Pb \cite{Mar98}, and $^{11}$Li+ $^{208}$Pb \cite{May99}. 
It is found that dipole polarizability gives rise to a significant reduction in the elastic scattering cross sections, 
which is particularly important for weakly bound nuclei.
For the case of $^{6}$He scattering on heavy targets, we expect that the 
strong Coulomb field 
 generated by the target can distort the $^6$He projectile, so that 
the $^{4}$He core is pushed away from the target nucleus, while the neutrons 
in the
halo remain unaffected.

  Another significant aspect is the behaviour of the nuclear optical potential. It
is an open question whether the optical model, using reasonable optical
potentials, is an adequate approach to describe the elastic scattering of
weakly bound nuclei. We are interested in determining the geometry and energy
dependence of the optical potential obtained from a fit to the elastic
cross sections, including the effect of dipole polarizability, and compare
it with the optical potentials obtained for the most similar stable nucleus,
which is $^{6}$Li.  The elastic scattering of $^{6}$He on $^{209}$Bi has been measured in Notre Dame
at energies near \cite{Kolataprl} and below \cite{Kolatarc} the Coulomb  
barrier. The analysis of the elastic scattering and reaction cross sections performed by  
the authors required the use of large, and energy-dependent, imaginary  
diffuseness parameters to reproduce the data.  
This result indicates the presence of a long range absorption mechanism,
which could be related to the effect of Coulomb excitation.

We have presented in a previous publication the analysis of elastic scattering of 
$^{6}$He on $^{208}$Pb at 27 MeV, measured at Louvain la Neuve 
\cite{Kakuee03}. We found evidence that an extremely large diffuseness 
parameter was required to fit the data. This indicates that long range 
reaction mechanisms occur when $^{6}$He was scattered on $^{208}$Pb. 
This is an important result, because it seems to indicate that the 
elastic scattering induced by exotic nuclei is qualitatively different from 
the scattering of stable nuclei. Nevertheless, it is important to confirm that
this feature (the long range absorption) occurs when  $^{6}$He collides with 
other heavy targets. If this is the case, we could recognise long range absorption as
a robust feature of the scattering of $^{6}$He at energies around the barrier, produced
by its weakly bound structure, that does not depend strongly on the target properties.

In this work we present new experimental data of the quasi-elastic scattering of
$^{6}$He on $^{197}$Au at 27 MeV. The $1/2^+$ state of $^{197}$Au at 70 keV was not resolved
from the $3/2^+$ ground state. We have explored the effect of including explicitly the 
excitation in a coupled channels calculation, and we find that the quasi-elastic differential 
cross sections (elastic plus inelastic) in the coupled channels calculation is very similar 
to the elastic differential cross section in an optical model calculation. 
The reason for it is that the flux going to the inelastic channels is subtracted from that 
of the elastic channel, leaving the quasi-elastic differential cross sections 
unaffected by the coupling.

We perform an analysis of the new set of data in parallel with
the analysis of the data of  $^{6}$He on $^{208}$Pb at the same energy. 
Our purpose is to search for evidence of long range absorption in 
these collisions.

\section{Experimental set-up}
  The experiment was performed using the radioactive beam facility at the
Cyclotron Research Centre at Louvain la Neuve in Belgium. The $^{6}$He
beam used in this experiment was produced via the $^{7}$Li(p,2p)$^{6}$He
reaction using LiF powder target contained in a graphite holder \cite{Vervier}.

The   post-accelerated secondary $^{6}$He beam of 27 MeV energy and intensity of
$3\times$10$^6$ ions/s was scattered on a  $^{197}$Au target, which  was in fact the backing of a $^{7}$LiF 
target. The thickness of the $^{197}$Au layer was approximately 300 $\mu g/cm^{2}$. 
The thickness of the $^{7}$LiF layer was
400 $\mu g/cm^{2}$.
The reaction products were  detected in a detection system
consisting of a LEDA and a LAMP type detector described in \cite{Tom2000}.
  LEDA and LAMP silicon strip detector arrays cover two different angular
ranges from 6$^\circ$-15$^\circ$ and from  23$^\circ$-72$^\circ$ in the laboratory frame, 
respectively. The details of the experimental setup has been described elsewhere \cite{Kakuee03}.

  Both the energy and the time of flight with respect to the cyclotron pulse
were recorded for each reaction product.  The in-beam energy resolution for
silicon strip detectors was around 120-140 keV, depending on the 
scattering angle, which is mainly due to the beam emittance and beam
straggling in the target.  The timing information in connection with energy
spectra enabled us to unambiguously identify the elastic scattering events.



\begin{figure}
\includegraphics[angle=270,width=0.8\textwidth]{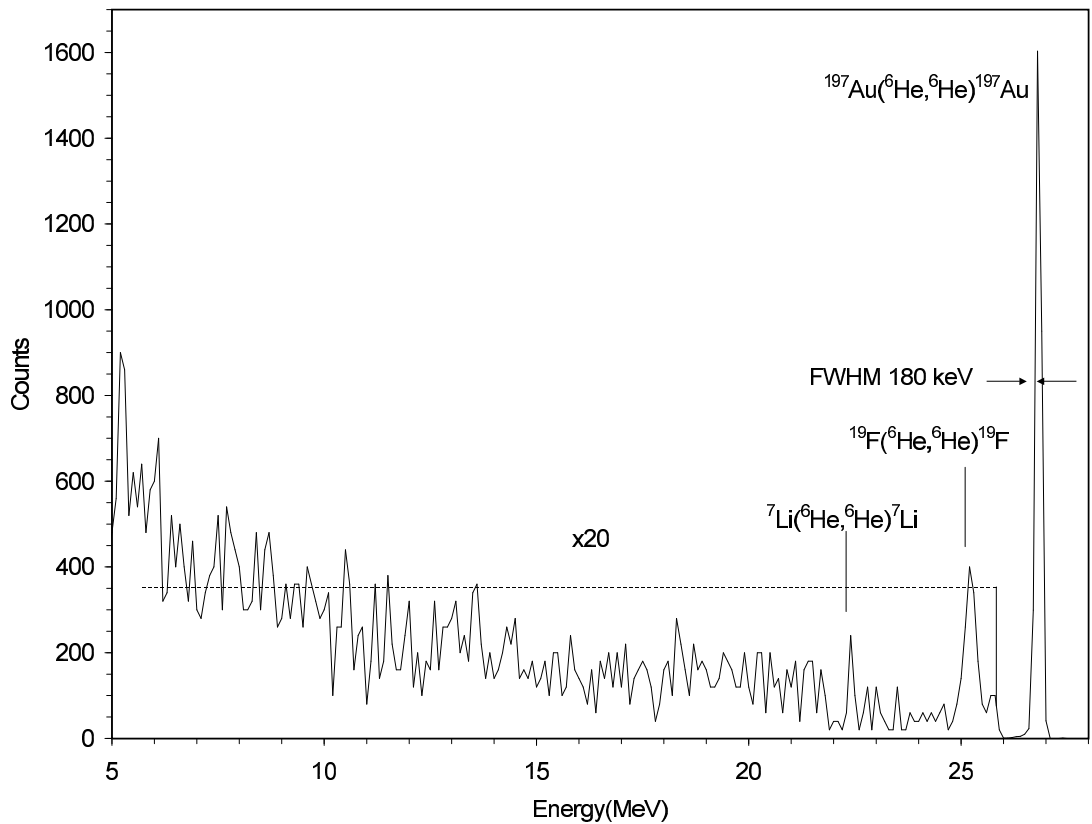}
\caption{The energy spectrum of 27 MeV  $^{6}$He scattered from  $^{7}$LiF(Au) at  $\theta_{lab}$=27$^\circ$}
\label{energy}
\end{figure}

The elastic scattering of
$^{6}$He from $^{197}$Au could be readily separated from $^{6}$He scattered from
$^{7}$Li and $^{19}$F at angles greater than 10 degrees. Figure \ref{energy} shows the separation of 
elastic scattering events corresponding to $^{197}$Au from those on $^{7}$Li and $^{19}$F.

There is no evidence of break-up into $^4$He 
at forward angles.
 At larger angles however there is some evidence for a broad break up
component around 2/3 of the elastic peak energy.
In the present paper no attempt has been made to extract information on break up events,
and only the elastic scattering events are investigated.

Only statistical uncertainties are considered in this analysis. 
The main source of systematic errors in our setup comes from possible 
uncertainties in the precise position of the LAMP array with respect to the 
target. This would lead to uncertainties in the scattering angle, which affect
the ratio of the measured cross sections to Rutherford cross sections.

In a previous work \cite{Kakuee03} we found that this uncertainty in the 
 position of LAMP affected mainly the relative normalisation of the 
small angles covered by LEDA and the large angles covered by LAMP. There, 
we found that minor changes in the positioning of LAMP ($\pm 3$ mm), affected 
the values of the long range absorption. As the small angles covered by LEDA
only give information on the global normalisation, and this normalisation, for
the intermediate angles (20 to 75 degrees) which are of interest, 
has systematic uncertainties, we
have chosen in this work to neglect the data at small scattering angles, and 
adjust the normalisation of the LAMP data to the theoretical calculations.

We have also considered the effect of adding the number of counts measured
in the different strips of the detector. Due to the geometry of LAMP array, we
have six strips which correspond to the same nominal scattering angle.
So, one could add the counts of all these strips, to reduce statistical 
uncertainties, as well as the number of data points. The result of applying 
this procedure is the ``averaged data'' set, shown in figure \ref{figcook}. 
This procedure is adequate
to visualise the trend of the data. However, we consider that this procedure 
could hide the presence of some systematic uncertainties, related to the 
different solid angles of the sectors of the LAMP array, as well as the 
different scattering angles corresponding to the strips in different sectors, 
due to beam misalignment. So, we have also considered the ``raw data'' set, in 
which one experimental data point is associated to each strip. In the
``raw data'' set, the effect of beam misalignment is taken into account 
\cite{Kakuee03},  
so that the scattering angles corresponding to strips of different sectors are
slightly different from the common nominal scattering angle.
The set of data
points so obtained, with more data and higher statistical uncertainties, are 
shown in figure \ref{figbestp}. 
The statistical significance of fits to the ``averaged data'' set would be the same than that of the ``raw data'' set, if the difference
in counting rates of the different sectors was purely statistical, and the difference between the actual scattering angle of each
strip and the common nominal scattering angle was negligible. As this hypothesis is not neccesarily true,
 we prefer to determine the optical potential parameters from a $\chi^2$ minimisation using the  ``raw data'' set, 
which is directly related to the experimental measurements.
However,  to
describe the qualitative features of the data is more adequate to use the
``averaged data'' of figure \ref{figcook}.

\section{Theoretical calculations}
  Dipole Coulomb excitation to break-up states can play an important role
in the scattering of weakly bound nuclei. To describe the effect of this
reaction mechanism in the elastic scattering, one can make use of a dynamic
polarisation potential \cite{May94}.
The form of the polarisation potential is obtained in a semi-classical
framework requiring that the second
order amplitude for the dipole excitation-deexcitation process and the first
order amplitude associated with the polarisation potential are equal for all
classical trajectories corresponding to a given scattering energy. This leads
to an analytic formula for the polarisation potential for a single excited state
\cite{May94}. The expression so obtained can be generalised for the case of
excitation to a continuum of break-up states \cite{May95} giving rise to the
following formula:
\begin{eqnarray}
\label{eq:Upol}
U_{pol}(r)&=&-\frac{4\pi }{9}\frac{Z_t^{2}e^{2}}{\hbar v}\frac{1}{(r-a_{o})^{2}r} \\
&&\int ^{\infty }_{\epsilon _{b}}d\epsilon \frac{dB(E1,\epsilon )}{d\epsilon }
\left( g(\frac{r}{a_{o}}-1,\xi )+if(\frac{r}{a_{o}}-1,\xi ),\right)  \nonumber
\end{eqnarray}
where \emph{g} and \emph{f} are analytic functions defined as
\begin{eqnarray}
f(z,\xi ) &=& 4\xi ^{2}z^{2} \exp(-\pi \xi )K_{2i\xi }''(2\xi z) \\
g(z,\xi ) &=& \frac{P}{\pi }\int _{-\infty }^{\infty }\frac{f(z,\xi ')}{\xi -\xi '}d\xi '
\end{eqnarray}
and \( \xi =\frac{\epsilon a_{o}}{\hbar v} \) is the Coulomb adiabaticity
parameter corresponding to the excitation energy \( \epsilon  \) of the nucleus.

When one compares the Coulomb dipole polarisation potentials for the 
collisions $^{6}$He + $^{208}$Pb, and $^{6}$He + $^{197}$Au, at the same 
energy of 27 MeV, one would expect that the larger charge of $^{208}$Pb 
would produce larger  dipole polarisation effects 
for $^{6}$He + $^{208}$Pb than for  $^{6}$He +$^{197}$Au  (note the factor
$Z_t^{2}$ in eq. \ref{eq:Upol}). However, the smaller charge of the $^{197}$Au 
target also makes the adiabaticity parameter $\xi$  smaller. 
This means that the imaginary part of the polarisation potential at long 
distances, describing long range absorption due to Coulomb break-up,  
is actually larger for $^{197}$Au than for $^{208}$Pb.

  The elastic differential cross sections are analysed assuming the 
validity of the optical model. The potential that describes the interaction  
between $^{6}$He and  
$^{208}$Pb is the sum of a monopole Coulomb potential, a dipole Coulomb  
polarisation potential and a phenomenological nuclear potential. The monopole 
Coulomb potential is determined by the charges of the colliding nuclei. 
Its only parameter is a Coulomb radius which, when taken in a reasonable 
range, does not affect significantly the cross sections. The dipole Coulomb  
polarisation potential describes the effect  of coupling the ground state to  
break-up states in the continuum by the dipole Coulomb force. 
It is a complex, long range and 
energy dependent potential, which is completely determined from the values 
of the B(E1) distribution of $^6$He. The phenomenological nuclear potential 
includes the ``direct'' term of the nuclear interaction between $^{6}$He 
and $^{208}$Pb as well as the dynamic effects of nuclear coupling to break-up 
states, quadrupole Coulomb coupling and Coulomb-nuclear interference terms. 
 
  For our analysis we take theoretical values of the B(E1) distribution of 
$^{6}$He \cite{ian,cobis}. This determines completely the dipole Coulomb 
polarisation potential.  The real part of the dipole polarization
potential is plotted as the dashed-dotted line in Fig. \ref{fig15X}, while the imaginary
polarization potential is plotted as the dashed line in Fig \ref{figwsaupb}.
For the nuclear potential, as a starting point, we use a Woods-Saxon 
potential, which was obtained 
from the optical model analysis of elastic scattering of $^{6}$Li, in mass 
range of 24-208 and 
energy range of 13-156 MeV \cite{Cook82}. We shall refer to this potential as
``Cook'' potential. The optical model parameters used are given in table 1.

 Our starting consideration is that $^6$Li and $^{6}$He have similar 
structures, 
and both are weakly bound (by 1.475 MeV and 0.975 MeV, respectively). So, the  
main qualitative difference between them could be that the dipole  Coulomb 
force can break up $^{6}$He into  $^{4}$He + 2n,  but it cannot break up  
$^{6}$Li  into $^{4}$He + $^{2}$H. The dipole Coulomb operator, in an 
$N=Z$ nucleus, is 
an isospin 1 operator. Since $^{6}$Li, $^{4}$He and $^{2}$H, have isospin 0 in 
their ground states,  it is not possible that the dipole Coulomb force breaks up $^{6}$Li into 
$^{4}$He + $^{2}$H.

\begin{figure}[tb]
\includegraphics[angle=0,width=\textwidth]{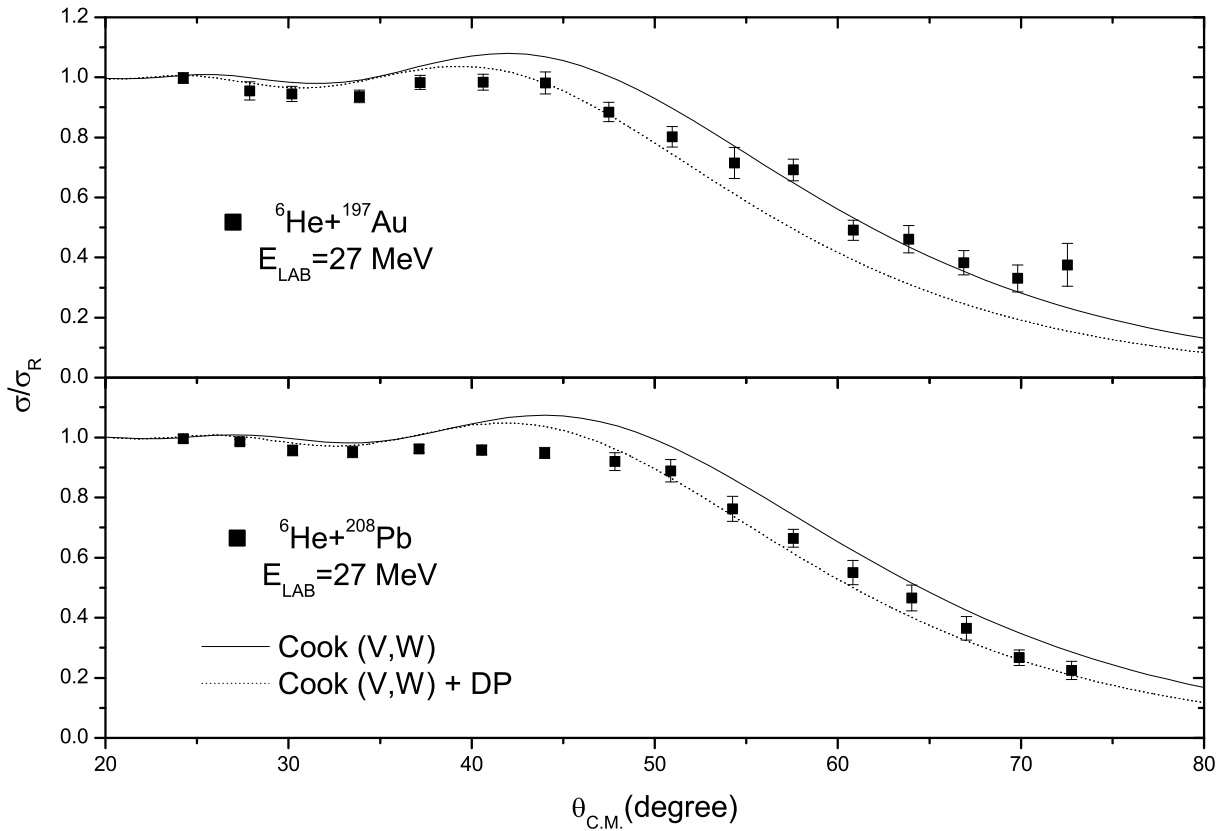}
\caption{Elastic scattering angular distribution of $^{6}$He + $^{197}$Au, $^{208}$Pb at 27 MeV. Data are ``averaged data'' (see text).
The lines are optical model calculations, without (solid line) and with (dashed line) taking into account 
dipole polarizability, which uses the same potential (real and imaginary) that fits $^{6}$Li systems 
(for details, see table \ref{tab1}).}  
\label{figcook}
\end{figure}

This difference between $^{6}$Li and $^{6}$He is explicitly taken into account by 
means of the dipole Coulomb polarisation potential. In figure \ref{figcook} 
we present the  ``averaged data'' measured for the collision of $^6$He on $^{197}$Au
and $^{208}$Pb at 27 MeV bombarding energy. The solid lines are the global optical model 
calculations which uses Cook potential \cite{Cook82}. 
These calculations, that do not take into account the effect of dipole 
polarizability, show
a well defined rainbow around 43 degrees, which is clearly absent from 
the experimental data. 
The calculations including dipole polarizability (dotted lines) also show 
a rainbow, but it is less pronounced and it appears at a 
smaller angle. So, we can conclude that the effect of dipole polarizability 
is clearly visible in the scattering of $^6$He on $^{197}$Au
and $^{208}$Pb at 27 MeV, and explains in part the disappearance of the rainbow
in the experimental data. However, it is also clear that the optical potential
obtained from  $^6$Li scattering (Cook potential), even supplemented with the
Dipole Polarisation Potential, is not adequate to reproduce the scattering of
  $^6$He on $^{197}$Au and $^{208}$Pb at 27 MeV.

\begin{table}[tb]
\begin{tabular}{|l|cccccccccc|} 
\hline
Projectile & model & V (MeV) & $W_{Pb}$ (MeV) & $a_i$ (fm) & $\chi^2$ & $\sigma_{R}^T (mb) $ & $\langle L \rangle_{T}$ & 
$\sigma_{R}^{dp} (mb)$ & $\langle L \rangle_{dp}$ & \\
\hline
$^6$Li & Cook  & 109.5 & 22.38 & 0.884 & --- & --- & --- & --- & --- &\\ 
$^6$He & Fit W, $a_i$ & 109.5 & 7.16 & 1.70 & 139 & 1934 & 15.7 & --- & --- &\\ 
$^6$He & DP ; Fit W, $a_i$ & 109.5 & 127.0 & 0.81 & 176 & 1713 & 14.9 & 302 & 26 & \\ 
$^6$He & Fit V, W, $a_i$ & 81.0 & 7.00 & 1.75 & 119 & 1970 & 16.2 & --- & --- & \\
$^6$He & DP ; Fit V, W, $a_i$ & 57.0 & 9.31 & 1.46 & 116 & 1878 & 16.2 & 318 & 24.9 &\\   
\hline
\end{tabular}
\caption{Optical model parameters obtained from $^{6}$Li + A (Cook systematics) and from 
$^{6}$He + $^{208}$Pb data analysis performed in this work. 
The radial parameters are fixed, for the real and the imaginary potential to 
$r_{0r}=r_{0i}=1.326$ fm 
and the real diffuseness parameter is fixed to $a_R=0.811$ fm. The number of data points is 82. 
$\sigma_{R}^T$ is the reaction cross section and $\sigma_{R}^{dp}$ represents the reaction cross section due only to the 
dipole polarizability.  $\langle L \rangle_{T}$ and $\langle L \rangle_{dp}$ are the respective average angular momentum 
weighted by the absorption cross section in each case.}
\label{tab1}
\end{table}

\begin{table}[tb]
\begin{tabular}{|l|cccccccccc|} 
\hline
Projectile & model & V (MeV) & $W_{Au}$ (MeV) & $a_i$ (fm) & $\chi^2$ & $\sigma_{R}^T (mb) $ & $\langle L \rangle_{T}$ & 
$\sigma_{R}^{dp} (mb)$ & $\langle L \rangle_{dp}$ & \\
\hline
$^6$Li & Cook  & 109.5  & 22.65 & 0.884 & --- & --- & --- & --- & --- & \\ 
$^6$He & {$a_i$($^{208}$Pb)}; Fit W,  & 109.5 & 6.20 & 1.70 & 135 & 1880 & 15.5 & --- & --- &\\ 
$^6$He & DP ; {$a_i$($^{208}$Pb)}; Fit W  & 109.5 & 84.3 & 0.81 & 141 & 1835 & 16.8 & 523 & 28.1 & \\ 
$^6$He & {V, $a_i$($^{208}$Pb)}; Fit W  & 81.0 & 6.12 & 1.75 & 113 & 1911 & 15.8 & --- & --- & \\
$^6$He & DP ; {V, $a_i$($^{208}$Pb)};  Fit W & 57.0 & 6.36 & 1.46 & 111 & 1907 & 17.6 & 564 & 26.6 &\\  
\hline
\end{tabular}
\caption{ Optical model parameters obtained from $^{6}$Li + A (Cook systematics) and from 
$^{6}$He + $^{208}$Pb data analysis according to table \ref{tab1} applied to $^{6}$He + $^{197}$Au 
angular distributions. The only fitted parameter is $W_{Au}$ which is the depth of the imaginary potential 
to fit $^{6}$He + $^{197}$Au data.  The number of data points is 80.}
\label{tab2}
\end{table}

\begin{figure}[tb]
\includegraphics[angle=0,width=0.8\textwidth]{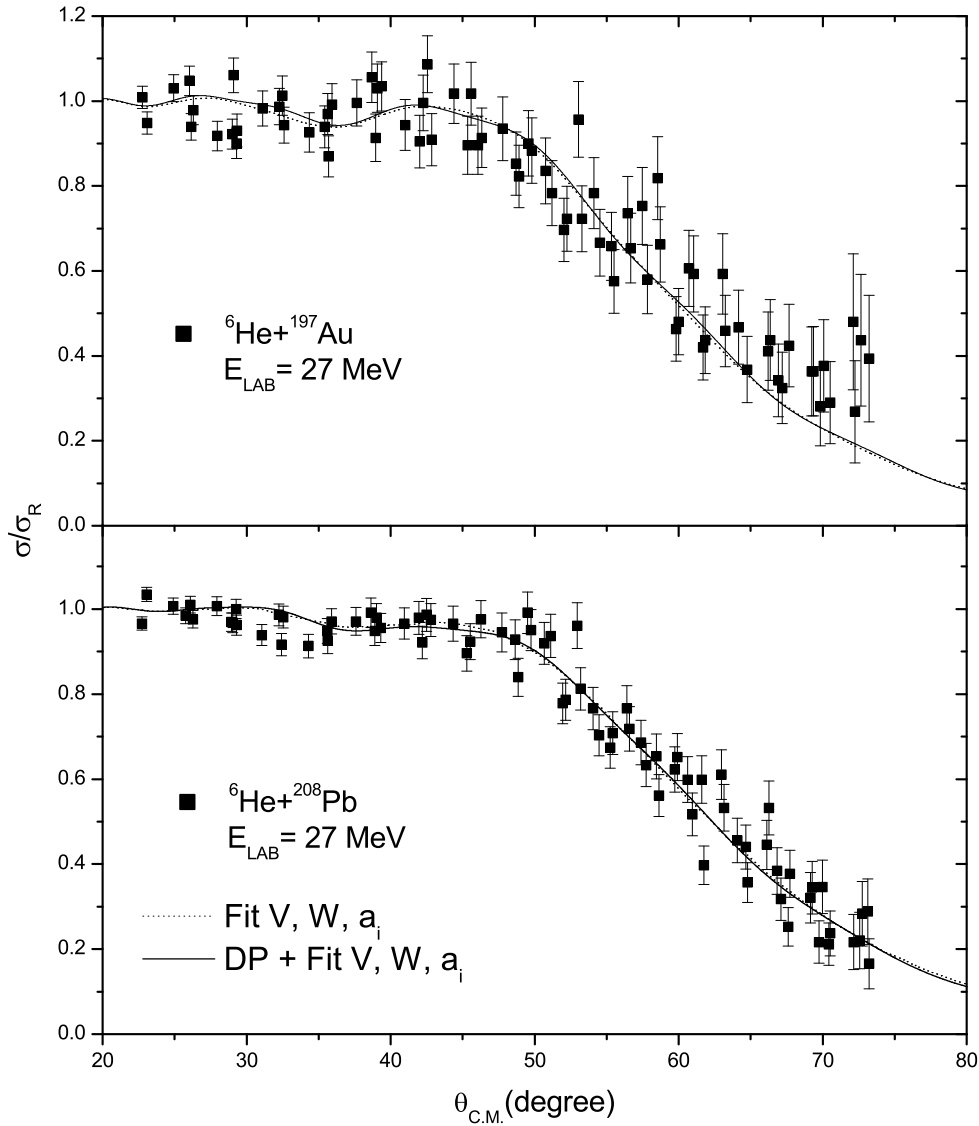}
\caption{Elastic scattering angular distribution of $^{6}$He + $^{197}$Au, $^{208}$Pb at 27 MeV. Data are ``raw data'' (see text).
The full line is an optical model calculation using Cook parameters with a 
reduced depth of real potential ($V=81 MeV$). The dashed line includes dipole polarizability 
effects and a reduced depth of real potential ($V=57 MeV$). The parameters ($W$, $a_{i}$) of the 
imaginary potential were also searched for a minimum $\chi^2$ in each calculation.}  
\label{figbestp}
\end{figure}

\begin{figure}[tb]
\includegraphics[angle=0,width=0.8\textwidth]{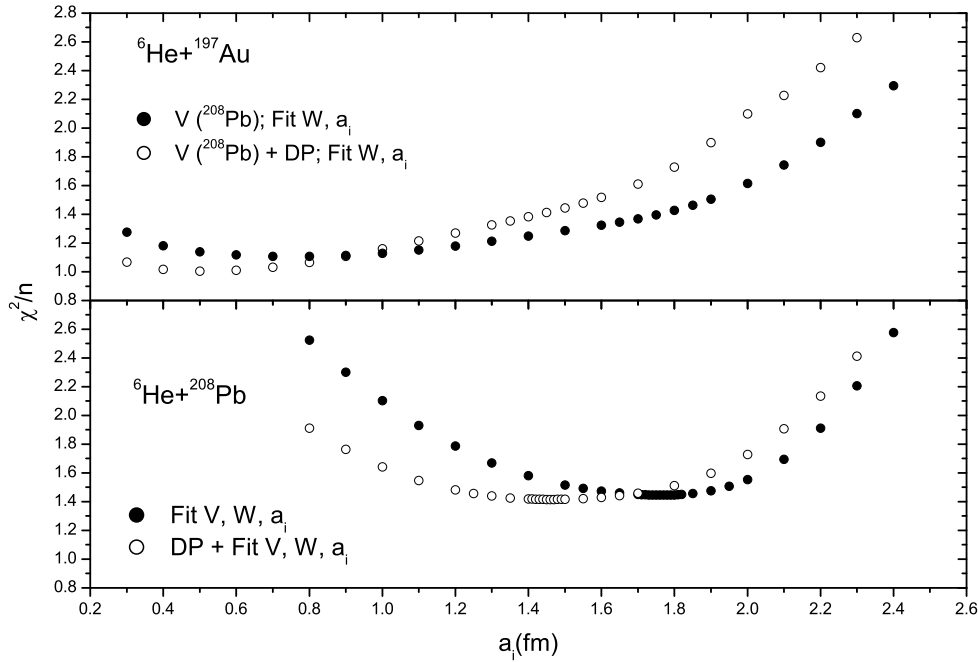}
\caption{Reduced $\chi^2$ for various values of imaginary depth and diffuseness ($W$, $a_{i}$) parameters, 
using the best real depth ($V$) without (closed symbols) and with (open symbols) the inclusion of the dipole 
polarisation potential, for the collision of $^{6}$He + $^{197}$Au, $^{208}$Pb.}  
\label{figchis}
\end{figure}

\begin{figure}[tb]
\includegraphics[angle=0,width=\textwidth]{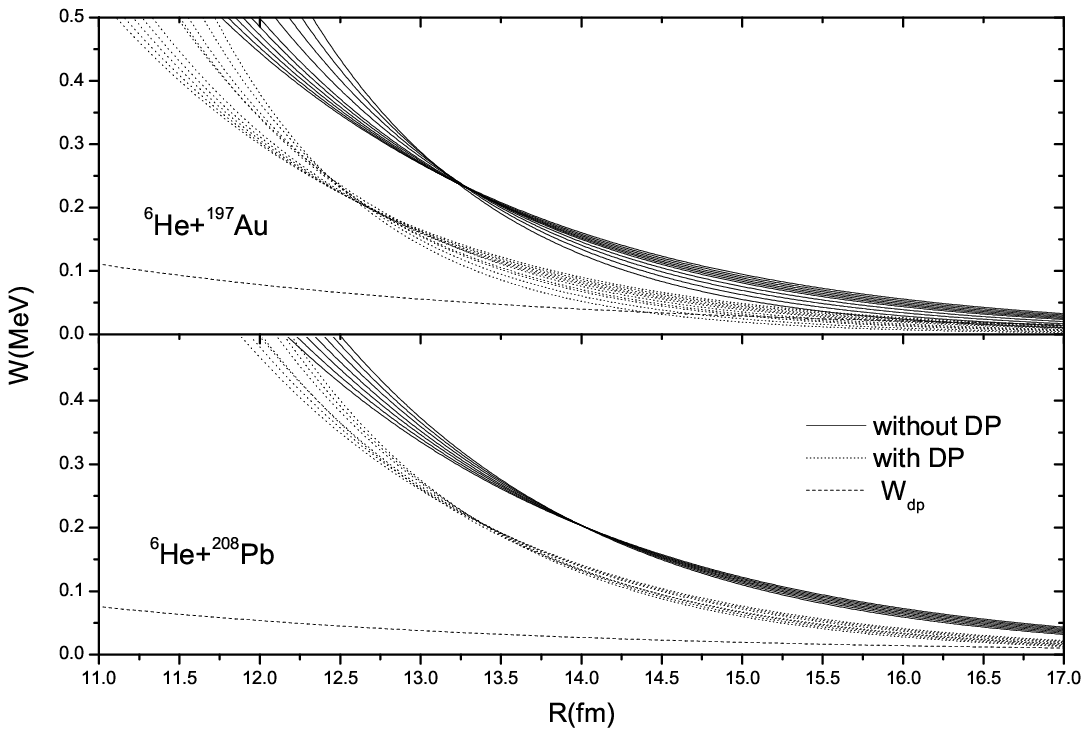}
\caption{Imaginary potentials that fit the data, including (dotted lines) 
and not including 
(solid lines) the dipole polarisation potential, as a function of the interacting distance. For the $^{6}$He + $^{208}$Pb case, we took $a_i=1.30 - 1.90$ fm.
For the $^{6}$He + $^{197}$Au case, we took $a_i=0.80 - 1.90$ fm. 
 The dashed line represent 
the  imaginary part of the dipole polarisation potential.}  
\label{figwsaupb}
\end{figure}

We now proceed to modify Cook $^{6}$Li optical potential to fit the $^{6}$He scattering data. 
The first argument is that the reaction channels
produced by  $^{6}$He scattering (mostly  break-up and neutron transfer), will
be different from those of  $^{6}$Li. These reaction channels  affect
mainly the imaginary part of the potential, which describes the loss of flux
from the elastic channel. So, we allowed the depth and the diffuseness of the 
potential to vary. The results of these calculations are shown in tables 
\ref{tab1} and \ref{tab2}. The fit of the data is fair, but not perfect, as
it can be seen from the values of $\chi^2$. If the value of the depth of the 
real potential is also fitted, then the fit of the data is very good, as it is
shown in tables \ref{tab1} and \ref{tab2}, as well as in figure 
\ref{figbestp}. 

In fitting the data, we have taken into account that the data
of  $^{6}$He on $^{208}$Pb are much more accurate than those of 
$^{6}$He on $^{197}$Au. So, whenever possible, we kept the same parameters for the two targets. 
We only allowed to vary the depth of the 
imaginary potential, because the targets $^{208}$Pb and $^{197}$Au could lead 
to different reaction channels. The parameters that fit the experimental data
of the reaction $^{6}$He on $^{197}$Au, shown in table \ref{tab2} are similar
to those obtained for the reaction $^{6}$He on $^{208}$Pb, shown in table 
\ref{tab1}. So are the values of $\chi^2$ obtained 
in the best fits. We can conclude that both sets of data give a consistent 
message, indicating the presence of long range reaction mechanisms.

It is interesting to comment on the values of the reaction cross sections 
obtained in the fits. These are around 1900 mb in all the calculations that
fit accurately the elastic data. From the reaction cross sections, we have 
extracted the contribution due to the
imaginary part of the Coulomb polarisation potential. This value is an 
estimate of the reaction cross section which is due to Coulomb break-up. Note
that this value is 318 mb for the $^{208}$Pb target, and 564 mb for the
 $^{197}$Au target. Naively, one would expect that the target with higher 
charge would induce more Coulomb break-up. However, the lower charge of 
 $^{197}$Au makes the energy of the collision (27 MeV) higher with
respect to the Coulomb barrier, reducing the collision time, and thus 
producing more Coulomb break-up. Also, it should be mentioned that Coulomb 
break-up leads indeed to long range absorption. This can be seen from the
values of $\langle L \rangle_{dp}$, which is the average L-value of the reaction cross 
sections due to coulomb excitation. They are considerably larger than
the values of $\langle L \rangle_{T}$, which is the average L-value of the total reaction 
cross sections.

 The results of the optical model fits discussed above, which have been performed with the code FRESCO 
\cite{FRESCO}, are shown in figure \ref{figbestp}, and the  sets of optical 
model parameters 
are shown in table \ref{tab1} ($^{208}$Pb) and table \ref{tab2} ($^{197}$Au). 
We have also performed calculations varying the value of $a_i$, and fitting the
value of $W$ for each $a_i$. The values of $\chi^2/n$ for these fits are 
plotted in Figure \ref{figchis}. The data on the $^{208}$Pb target indicate
clearly the need for large imaginary diffuseness parameters, to obtain good 
fits $\chi^2/n \le 1.5$. However, the data on the $^{197}$Au target can be 
fitted with the same accuracy with almost any diffuseness parameter. This is
an effect of the larger statistical uncertainties of the  $^{197}$Au data. 
In any case, we can say that the large imaginary diffuseness parameter 
required to reproduce the data on the $^{208}$Pb target is not inconsistent 
with the values required to reproduce the data on the $^{197}$Au target. 
Also, the fits show that the explicit inclusion of the Dipole Polarisation 
Potential reduce the values of the diffuseness parameters required to fit 
the data in both cases.


The fits presented in tables \ref{tab1} and \ref{tab2}, and in figure \ref{figchis} make use of the ``raw data'' set, 
which are shown in figure \ref{figbestp}. We have also performed the optical potential fits making use of the ``averaged data''
set, presented in figure \ref{figcook}. We find that the results for the potentials that produce the best fits are very similar 
in both cases. This indicates that, in these experiments, there were not systematic differences between the cross sections obtained 
from the detector strips of the different sectors, and that the difference between actual and nominal scattering angles was not 
important, for the observable considered.

\begin{figure}[tb]
\includegraphics[angle=0,width=0.8\textwidth]{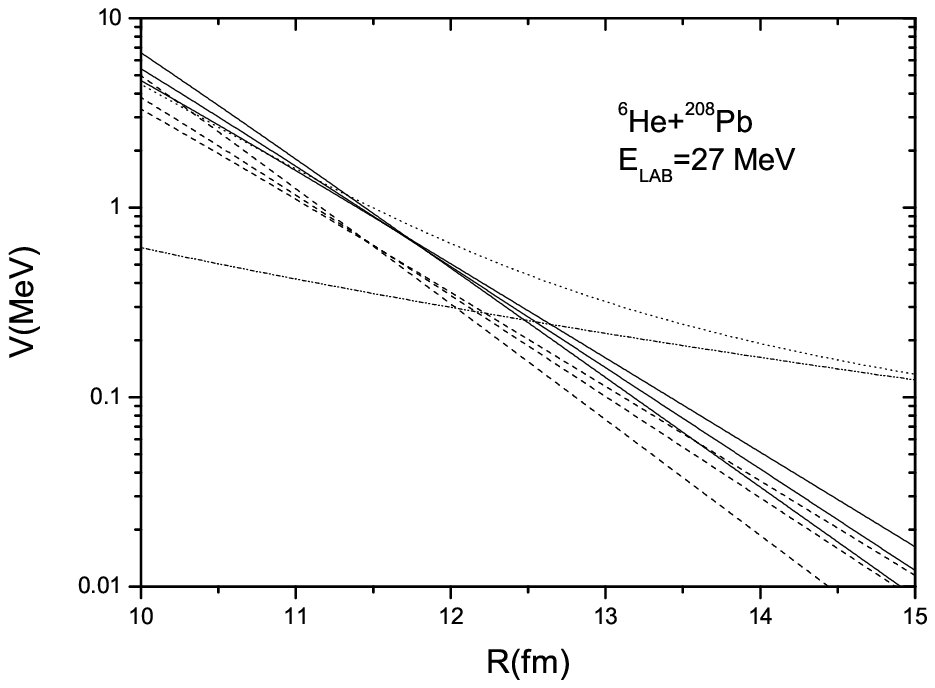}
\caption{Real potentials that fit the data including (dashed lines) 
and not including 
(solid lines) the dipole polarisation potential, as a function of the interacting distance. The central line is the best fit, and the other lines are obtained
varying slightly the diffuseness parameter, and readjusting the potential depth.
The dashed-dotted line is the real part of the DPP due to dipole polarizability. The dotted line is the sum of the real potential, including DPP, 
plus the real part of the DPP. 
This calculation agrees with the calculations performed without including explicitly a DPP (solid lines),  to the strong absorption radius $R_{SA}=11.5$ fm}  
\label{fig15X}
\end{figure}

\section{Discussion}

One objective of this analysis  is to investigate if, as suggested in our 
previous paper \cite{Kakuee03}, there are evidences of long range mechanisms 
that lead to the loss of flux in the elastic channel at kinematic conditions 
that suggest the nuclei are far beyond the strong absorption radius, which, 
in our case, is approximately $R_{SA}=11.5 fm$ for both targets.
 
The other objective is to investigate the role of coulomb dipole 
polarizability in the scattering of this weakly bound, and hence 
easily polarizable, nucleus.

\subsection{Evidence for long range absorption mechanisms}

The first evidence for the long range absorption comes from the values of
the imaginary diffuseness parameters required to fit the data.
The plot of $\chi^2$ (figure \ref{figchis}), for the $^{6}$He + $^{208}$Pb, 
clearly indicates that the imaginary potential needed to fit the data has a diffuseness considerably larger 
than that of the real potential. A diffuseness parameter in the range of 1.30 to 
1.90 fm is needed to fit the 
data. This is to be compared to the value of the diffuseness of the real 
potential, which is $a_r = 0.811$ fm.
We point out that this diffuseness is not as large as the one found in the 
previous work \cite{Kakuee03}, but 
there the diffuseness of the real potential was allowed to vary, and it also 
acquired large values. The systematics of our calculations show that the 
imaginary potential has to be much more diffuse than the real potential to
reproduce the data.

The second evidence for the long range absorption comes from the relatively 
large values of the average reaction angular momenta $<L>_T$. In a cutoff
model, which is not unreasonable to describe the absorption in heavy ion 
collisions, the absorption is maximum for $L+1/2<\lambda$ and negligible
for $L+1/2 > \lambda$. The relation of the cutoff parameter $\lambda$ and the
reaction cross section is given by \cite{satchler} 
$\sigma_R = \pi \lambda^2/ k^2$. Then, in this sharp cutoff model, 
$\langle L \rangle_T+1/2=2/3 \lambda$. The values of the reaction cross sections in these 
reactions (see tables \ref{tab1},\ref{tab2}) are about 1900 mb. This leads to  values
of $\langle L \rangle_T \simeq 13.5$, which is considerably smaller than the values shown
in the table. This is an indication of the fact that the reaction cross 
sections extend to up to values of $L$ which are well beyond the grazing 
angular momentum.

The third evidence for the long range absorption comes from the values of the
imaginary potential. In figure \ref{figwsaupb} different families of imaginary 
potentials which fit reasonably the data are plotted as a function of the 
distance. Let us focus
on the solid lines, which correspond to optical model calculations
in which the dipole polarisation potential (DPP) is not explicitly included.
We see that the lines cross at 13.25 fm (Au) and 14.0 fm (Pb). 
This indicates the region of
sensitivity to the imaginary potential. Note also that the imaginary potential 
still has sizeable values at distances as large as 16 fm. In 
figure \ref{fig15X} we present real potentials that fit the data. These 
potentials cross between 11 and 12 fm, which is the region of the strong 
absorption radius (11.5 fm). So, we see that while the real potential is 
determined around the distance of the strong absorption radius, the 
imaginary potential is determined at much larger distances. This point should 
be taken into account when investigating the energy dependence of the real 
and imaginary optical potentials of exotic nuclei.

It should be noticed that these evidences for long range absorption are deduced from a 
consistent analysis of the scattering data on $^{197}$Au and $^{208}$Pb targets.
The data on the $^{197}$Au target, considered separately, 
are not sufficiently accurate to determine unambiguously the depth and diffuseness of the 
 imaginary potential. However, the other two evidences for long range absorption
 (large values in the average reaction angular momenta, and
sensitivity of the imaginary potential to large distances) come out clearly
from the analysis of the $^{197}$Au data. These features are unaffected by the ambiguities
of the imaginary potential parameters, as  shown in figure \ref{figwsaupb}.

\subsection{Role of Coulomb dipole polarizability}

Having established the presence of long range reaction mechanisms, we will now
discuss the relevance of the Coulomb dipole polarizability in this mechanism.
For that purpose, we will consider the calculations in which the  DPP is 
explicitly included. In these calculations, the phenomenological imaginary 
potential describes the absorption produced by dynamical effects different from
pure dipole Coulomb excitation.

First, we investigate the change in the values of the imaginary diffuseness 
parameters. As shown in table \ref{tab1}, the value of the diffuseness parameter
when the DPP is explicitly included is 1.46 fm, to be compared with 1.75 fm when it
was omitted. This can also be seen in figure \ref{figchis}, where the values 
of the diffuseness parameters which reduce the values of $\chi^2$ are 
definitively smaller when the DPP is included. Note that in the $\chi^2$ plot 
for $^6$He + $^{197}$Au, the relevant magnitude is not the absolute minima.
In the case of  $^6$He + $^{208}$Pb, the best fits obtained had
$\chi^2/n \simeq 1.5$. We can argue that, for the less accurate 
 $^6$He + $^{197}$Au data, a fit with
$\chi^2/n \le 1.5$ already indicates a good fit of the data. Hence, the plots
should be interpreted saying that, when the DPP is not explicitly included, the
imaginary diffuseness parameter should be less than 1.90 fm, while if the
DPP is explicitly included, it should be less than 1.60 fm. This is compatible 
with the values obtained with the $^6$He + $^{208}$Pb data. 
The data therefore indicate that when the DPP is  included, there is
a need of long range imaginary potentials, but the range is not as large as 
when the DPP is not  included.

Second, we investigate the values of the reaction cross sections and average 
L-values produced by the DPP. As shown in tables \ref{tab1}, \ref{tab2}, a 
significant
fraction of the reaction cross section is due to the coulomb dipole excitation
mechanism. In addition, the values of the average angular momenta are 
very large
(24.4 and 26.6 respectively for the $^{197}$Au and $^{208}$Pb targets). 
These values of the angular momenta correspond to distances of closest approach
of 14.6 and 14.8 fm, which are well beyond the strong absorption radius.
This indicates that coulomb dipole polarizability is indeed a mechanism that 
generates long range absorption. This produces a reduction of the elastic 
cross sections at forward angles, which is associated with the disappearance of
the rainbow previously discussed.
It should be noticed that the absorption cross section due to the effect of 
dipole polarizability does not increase with the charge of the target. This is
due to the effect of the adiabaticity parameter $\xi$, which is larger for the
$^{208}$Pb target than for the $^{197}$Au target. This means that, although the
coulomb dipole force is weaker in the  $^{197}$Au target, it is more effective 
in producing break-up cross sections, which generate absorption in the elastic
channel. 

Third, we investigate the values of the potentials as a function of the 
distance. Consider the dashed lines in figure \ref{figwsaupb}. They represent
the imaginary potentials which describe absorption by mechanisms different 
from the dipole polarizability. They cross at distances of 12.5 (Au) and 
13.25 (Pb) fm, which are considerably smaller than the crossings of the full
lines, where the DPP is not explicitly considered. So, we see explicitly that
a long range absorption mechanism is required in addition to
 the pure dipole coulomb excitation. However, the range of the imaginary 
potential associated to this additional mechanism
is not as large as the imaginary DPP, which is shown by the dotted line.

The DPP has a real component, attractive, which is shown by the dot-dashed 
line. This potential has a very long range, but its effect on the scattering 
is determined mainly by its value at the strong absorption radius. As shown in
figure \ref{fig15X}, all the calculations that fit the data, either with or 
without the explicit inclusion of the DPP, have values of the real potential
which are about 0.90 MeV at 11.5 fm. However, the long range attraction 
 enhances the absorption by the imaginary potentials. This
explains that the sum of the DPP imaginary potential and the phenomenologic
potentials given by the dashed lines in figure \ref{figwsaupb} are smaller than
the full lines, which represent the phenomenologic imaginary potentials where
DPP is not considered.

We can conclude that the effect Coulomb Dipole Polarizability accounts for an
important fraction of the long range absorption needed to describe the
elastic scattering of $^{6}$He on $^{208}$Pb and  $^{197}$Au at 27 MeV.
However, other reaction mechanisms such as nuclear break-up, coulomb-nuclear
interference effects or neutron transfer to bound or unbound states can play
also an important role.

We consider that a proper understanding of the long range absorption is 
required. The use of nuclear reactions as an spectroscopic tool to 
investigate the structure of exotic nuclei requires a deep 
understanding of the reactions induced by exotic nuclei. This work indicates
that simple preconceptions based on the experience of the optical model
on stable nuclei, such as the role of the strong absorption radius, are not
extrapolatable to the scattering of exotic nuclei. 

Experimental measurements of the elastic scattering of $^6$He and 
other exotic nuclei on a variety of targets, along with the measurement of the
main reaction channels, would be required. Reaction calculations with a proper
treatment of the break-up channels would be needed to understand the role of 
absorption.

\bigskip

{\bf Acknowledgements:}

The authors thank the staff at the Cyclotron Research Centre 
accelerator
facility in Louvain-la-Neuve, Belgium, for providing us with an intense and
good quality radioactive beam. 
ORK acknowledges a grant from the Iranian government and support from the 
University of Sevilla for his stay at this university.
PJW, TD, ACS, AL, ADP and WBS would like to acknowledge support from the British EPSRC.
AMSB, IMB, MVA, AM, MAGA and JGC would like to acknowledge support from the Spanish MCyT
under projects  FPA2003-05958 and FPA2005-04460. AMSB
acknowledges a research grant from the Spanish MCyT.


\end{document}